\documentstyle[graphics]{mn}

\newif\ifAMStwofonts
\ifoldfss
  \ifCUPmtlplainloaded \else
    \NewTextAlphabet{textbfit} {cmbxti10} {}
    \NewTextAlphabet{textbfss} {cmssbx10} {}
    \NewMathAlphabet{mathbfit} {cmbxti10} {} 
    \NewMathAlphabet{mathbfss} {cmssbx10} {} 
  \fi
  \ifAMStwofonts
    \ifCUPmtlplainloaded \else
      \NewSymbolFont{upmath} {eurm10}
      \NewSymbolFont{AMSa} {msam10}
      \NewMathSymbol{\upi}     {0}{upmath}{19}
      \NewMathSymbol{\umu}     {0}{upmath}{16}
      \NewMathSymbol{\upartial}{0}{upmath}{40}
      \NewMathSymbol{\leqslant}{3}{AMSa}{36}
      \NewMathSymbol{\geqslant}{3}{AMSa}{3E}

    \fi
  \fi
\fi 

\ifnfssone
  \newmathalphabet{\mathit}
  \addtoversion{normal}{\mathit}{cmr}{m}{it}
  \addtoversion{bold}{\mathit}{cmr}{bx}{it}
  \newmathalphabet{\mathbfit} 
  \addtoversion{normal}{\mathbfit}{cmr}{bx}{it}
  \addtoversion{bold}{\mathbfit}{cmr}{bx}{it}
  \newmathalphabet{\mathbfss} 
  \addtoversion{normal}{\mathbfss}{cmss}{bx}{n}
  \addtoversion{bold}{\mathbfss}{cmss}{bx}{n}
  \ifAMStwofonts
    \ifCUPmtlplainloaded \else
      \UseAMStwoboldmath
      \makeatletter
      \new@mathgroup\upmath@group
      \define@mathgroup\mv@normal\upmath@group{eur}{m}{n}
      \define@mathgroup\mv@bold\upmath@group{eur}{b}{n}
      \edef\UPM{\hexnumber\upmath@group}
      \new@mathgroup\amsa@group
      \define@mathgroup\mv@normal\amsa@group{msa}{m}{n}
      \define@mathgroup\mv@bold\amsa@group{msa}{m}{n}
      \edef\AMSa{\hexnumber\amsa@group}
      \makeatother
      \mathchardef\upi="0\UPM19
      \mathchardef\umu="0\UPM16
      \mathchardef\upartial="0\UPM40
      \mathchardef\leqslant="3\AMSa36
      \mathchardef\geqslant="3\AMSa3E
    \fi
  \fi
\fi 

\ifnfsstwo
  \DeclareMathAlphabet{\mathbfit}{OT1}{cmr}{bx}{it}
  \SetMathAlphabet\mathbfit{bold}{OT1}{cmr}{bx}{it}
  \DeclareMathAlphabet{\mathbfss}{OT1}{cmss}{bx}{n}
  \SetMathAlphabet\mathbfss{bold}{OT1}{cmss}{bx}{n}
  \ifAMStwofonts
    \ifCUPmtlplainloaded \else
      \DeclareSymbolFont{UPM}{U}{eur}{m}{n}
      \SetSymbolFont{UPM}{bold}{U}{eur}{b}{n}
      \DeclareSymbolFont{AMSa}{U}{msa}{m}{n}
      \DeclareMathSymbol{\upi}{0}{UPM}{"19}
      \DeclareMathSymbol{\umu}{0}{UPM}{"16}
      \DeclareMathSymbol{\upartial}{0}{UPM}{"40}
      \DeclareMathSymbol{\leqslant}{3}{AMSa}{"36}
      \DeclareMathSymbol{\geqslant}{3}{AMSa}{"3E}
    \fi
  \fi
\fi 

\ifCUPmtlplainloaded \else
  \ifAMStwofonts \else 
    \def\upi{\pi}
    \def\umu{\mu}
    \def\upartial{\partial}
  \fi
\fi

\def\kms{km s$^{-1}$}
\def\etal{{et al.\ }}
\def\eg{{\it e.g.\ }}
\def\ie{{\it i.e.\ }}
\def\solmtxt{M$_{\odot}$}

\title{Gas Inflow in Barred Galaxies --- Effects of Secondary Bars}
\author[Witold Maciejewski, Peter J. Teuben, Linda S. Sparke and James M. Stone]
{Witold Maciejewski$^{1,2}$, Peter J. Teuben$^{3}$, Linda S. Sparke$^{4}$ and James M. Stone$^{3}$\\
$^1$Theoretical Physics, 1 Keble Road, University of Oxford, Oxford, OX1 3NP,
{\tt witold@thphys.ox.ac.uk}\\
$^2$Jagiellonian University Observatory, ul. Orla 171, Krak{\'o}w, Poland\\
$^3$Department of Astronomy, University of Maryland, College Park, MD 20742\\
$^4$Department of Astronomy, University of Wisconsin, 475 N Charter St, Madison, WI 53706}

\begin{document}

\maketitle

\begin{abstract}
We report here results of high-resolution hydrodynamical simulations
of gas flows in barred galaxies, with a focus on gas dynamics in the
central kiloparsec. In a single bar with an Inner Lindblad 
Resonance, we find either near-circular 
motion of gas in the nuclear ring, or a spiral shock extending towards
the galaxy center, depending on the sound speed in the gas. From a simple 
model of a dynamically-possible doubly barred galaxy with resonant coupling, 
we infer that the secondary bar is likely to end well inside its corotation.
Such a bar cannot create shocks in the gas flow, and therefore will not
reveal itself in color maps through straight dust lanes: the gas flows 
induced by it are different from those caused by the rapidly rotating 
main bars. In particular, we find that secondary stellar bars are unlikely to
increase the mass inflow rate into the galactic nucleus.
\end{abstract}


\section{Introduction}
Gas in disc galaxies develops sharp and well-defined characteristic
morphological features, as a response of this dynamically cold and 
dissipative medium to the structural details of the underlying gravitational
potential. In barred galaxies, the most distinctive of these features is
a special category of dust lanes, where the gas and dust get compressed. 
These often appear as two symmetric, almost straight lanes, slightly 
tilted to the bar major axis, and located at the leading edge of the bar 
(Athanassoula 1984). Throughout this paper, we will call them the {\it 
principal dust lanes}. Another prominent feature is a set of resonance 
rings (see e.g. Buta 1986, Buta 1995), which take oval or circular shapes. 

Both the theory of orbital structure, and hydrodynamical modeling of gas in 
barred galaxies, provide a consistent picture that relates
the characteristic gas features to the galaxy's dynamics (see 
Sellwood \& Wilkinson 1993 for a review). If the bar has an Inner Lindblad 
Resonance (ILR), gas can populate the two major families of stable orbits 
inside the bar: the $x_1$ orbits
elongated with the bar, and $x_2$ orbits perpendicular to it. Around the
radius of the ILR, orbiting gas shifts gradually from the $x_1$ orbits 
to the $x_2$ orbits, creating two {\it principal shocks} along the leading 
edges of the bar (Athanassoula 1992). Gas and dust are compressed by the 
shock, causing large extinction, so we see these regions as the
dark {\it principal dust lanes} in the galaxy image.
The shocked gas loses angular momentum to the 
bar --- hydrodynamical models indicate a considerable gas inflow towards 
the center. If gas follows closely the periodic orbits in the bar
(is dynamically cold), it 
should eventually settle on the nuclear ring (Piner, Stone \& Teuben 1995; 
Patsis \& Athanassoula 2000), located approximately at the position of 
the bar's ILR. By contrast to the principal dust lanes, gas concentrated in 
the nuclear ring experiences little shear, and star formation is likely 
to occur there, causing the blue appearance of this ring (Buta 1986). 

In this picture, bars provide a mechanism for moving gas from the 
galactic disk in towards the centre to create a starburst.
However, Active Galactic Nuclei (AGN) require mass accretion to 
extend to within a few parsecs of the galaxy center, much closer in
than the usual position of bar's ILR, which usually has a radius at
least 10\% of that at corotation.
If orbits of the gas flow oscillate more freely around 
the periodic orbits (gas is dynamically warmer), such flow can
be driven closer to the galactic center (Englmaier \& Gerhard 1997,
Patsis \& Athanassoula 2000). 

Additionally, a mechanism of bars within bars 
has been invoked to drive further inflow, and thus feed the AGN in 
a manner similar to the inflow on large scales (Shlosman, Frank \& 
Begelman 1989). It involves a cascade of nested, independently rotating 
bars, that ensure continuous gas inflow to the very central regions.
Arguments for the existence of multiple (or at least double) bars
come from both theory and observations. They emerge as a rather
common phenomenon, from a survey of 38 nearby early-type 
barred field galaxies (Erwin \& Sparke 1999), which indicates that 
at least $\sim$20\% of them possess secondary bars. The inner bar 
characteristically is 5 to 7 times smaller than the main one. It is 
oriented randomly with respect to the main bar (Buta \& Crocker 1993,
Friedli 1996),
which one should expect if they are dynamically distinct subsystems.
The presence of secondary bars in infrared images, indicates that they
are not exclusively made out of gas, but rather contain old stars,
and thus form independent systems from the point of view of stellar
dynamics. The reason for this may be that orbital times in the inner galaxy 
are much smaller than those at radii of a few kiloparsecs, and the
dynamically decoupled inner bar should rotate faster than the outer 
structure. Simple analytical models (Maciejewski \& Sparke 1997)
indicate that periodic feeding of the nucleus may occur in this scenario.

To test this hypothesis, one can follow gas dynamics in combined 
gas and stars numerical simulations in which such double bars form 
(\eg Friedli \& Martinet 1993, who used Smoothed Particle Hydrodynamics,
\ie the SPH algorithm). Nevertheless, 
the resolution and accuracy achieved so far is rather low. On the other 
hand, one can assume potentials and pattern speeds of the two stellar bars, 
and study the gas response (Athanassoula 
2000). This paper follows the second approach, but instead of choosing the
parameters of the bars arbitrarily, we take them from a dynamically possible 
model of a doubly barred galaxy (Maciejewski \& Sparke 2000, hereafter Paper 
I), which was constructed by searching for stellar orbits that can support a
potential of two nested bars with different rotation periods.

Numerical models of gas flow in barred galaxies can be traced back
more than twenty years (see Athanassoula 1992, and Sellwood \& Wilkinson 
1993 for reviews), but an overwhelming increase in computing speed
and memory makes the recent models considerably more detailed, and
inclusive of a greater variety of physical phenomena. 
Despite this progress, the rate of inflow is still not well determined,
with methods based on SPH algorithms predicting inflow that is orders of 
magnitude higher than that in grid-based methods (compare Piner \etal 1995 
with Patsis \& Athanassoula 2000). A landmark in this field
is a set of about two hundred hydrodynamical models, created by Athanassoula 
(1992) with an Eulerian code. These explore the dependence
of the gas flow on various parameters of the model potential.
By selecting models that reproduce the observed shape of the
principal dust lanes, Athanassoula showed that bar's corotation
should be located just past the end of the bar. One can also infer 
from this set of models that an ILR is necessary for the principal
dust lanes to be offset from the center, as seen in many real galaxies.

Piner \etal (1995) repeated some of these calculations with better
resolution on a polar grid. They noticed that if an ILR is
present, a nuclear ring forms inside this resonance.
Virtually no gas passes inside the nuclear ring in their simulation
to reach the nucleus. If the ILR is absent the nuclear ring does not
form, and strong inflow to the center can take place, as found by
Athanassoula (1992), and confirmed by Piner \etal (1995).
Englmaier \& Gerhard (1997) noticed that the principal shock 
structure depends on the gas sound speed: the off-axis shock
at low sound speed turns into an on-axis one when the sound 
speed is high. Recently Englmaier \& Shlosman (2000)
have used a grid-based algorithm in polar coordinates to study gas
flows in central parts of barred galaxies.

We use high resolution hydrodynamical simulations to examine in detail
the gas flow in barred galaxies, with particular interest in the flows 
within the inner 1 kpc. In \S2, we present our code, and describe the model.
In \S3, we revise and extend the basic picture of gas flow in a bar,
that we outlined above. In \S4, we focus our attention on the inner 1 kpc, 
and explore the  nature of the nuclear ring, and gas flow in the presence 
of a dynamically possible inner secondary bar. We find that although it is 
not likely that the secondary bar enhances inflow in the inner galaxy, 
there are other conditions permitting strong inflow. We relate our models 
to observations in \S5.

\section{Numerical modeling of gas flows in galaxies}
It is difficult to model the interstellar medium (ISM) in galaxies
on global scales, because it has a complex morphology that can be
traced to sizes many orders of magnitude smaller than the scale of interest
(Sellwood \& Wilkinson 1993). The gaseous ISM consists of several phases 
with considerably different physical properties, and is highly turbulent,
and probably fractal (see \eg Elmegreen 1997, Padoan, Jones \& Nordlund
1997). If a single-fluid approximation is used
to model global gas features, this fluid has no commonly accepted
equation of state. For this reason a variety of numerical methods has been
developed for the study of global ISM dynamics in galaxies --- we refer the 
reader to Teuben (1995) for a review. In this paper, we implement
the full solution of the hydrodynamical equations in the Eulerian
formulation on a fixed grid.

\subsection{Our code, simulation properties, and initial conditions}
We used the grid-based Eulerian code CMHOG, written by James M. Stone 
at the University of Maryland, and briefly described by Piner \etal (1995).
It solves the single-fluid hydrodynamic equations in their conservative
form, i.e., the continuity equation
\begin{equation}
\frac{\partial \rho}{\partial t} + \nabla \cdot (\rho {\bf v}) = 0 , \\
\end{equation}
and Euler's equation
\begin{equation}
\frac{\partial (\rho \bf{v})}{\partial t} + 
\nabla \cdot (\rho {\bf v} {\bf v}) = - \nabla P - \rho \nabla \Phi ,
\end{equation}
where the ${\bf v} {\bf v}$ in the second term represents the
velocity tensor $T_{ij} = v_{i} v_{j}$. 
The third (energy) hydrodynamic equation is replaced by assuming an 
isothermal equation of state, which closes the system of 
equations defining the model. The typical cooling time in the
problem is much shorter than the time-step, and the isothermal equation 
of state forces the gas to cool instantly. The gas temperature is
fixed to model the statistical behaviour of gas clouds. Self-gravity
of the gas is not taken into account.

The CMHOG code solves the hydrodynamical equations above by implementing 
the PPM (piecewise parabolic method) algorithm (Woodward \& Colella 1984)
in its Lagrangian remap formulation. The only viscosity in this code 
is the numerical one, and a test
with gas in circular motion has been performed to show that it is
too small to noticeably alter the flow within times considered in 
this work.

The most appropriate grid for disk modeling is a polar one, and Piner \etal 
(1995) reformulated the CMHOG code to use planar polar coordinates in 2D, 
which we adopt here. Because the PPM algorithm works best for nearly square 
grid cells ($\Delta R \simeq R\Delta\varphi$), and since the number of 
tangential zones is fixed, the cell size is proportional to $R$.
The resolution near the grid center is excellent, so the polar grid 
is suitable to accurate study of circumnuclear phenomena in galaxies. 
On the other hand, the time-step, which is related to the cell size by
the Courant condition, becomes prohibitively short unless the inner 
boundary is imposed at a reasonable distance from the grid center.
This is a disadvantage of a polar 
grid, because it requires additional boundary conditions and places
the innermost galaxy outside the modeled domain. 

Calculations for all our models were performed in two dimensions 
on polar grids extending
outwards to 16 kpc in radius. We do not extend our grids inwards 
beyond the inner 20 pc (the extended grids Le and Me), and cut them
at 100 pc when possible (grids Ls, Ms, and Hs). Grids used in this work 
are listed in Table 1. In order to achieve the resolution of the Me grid 
near the center, a Cartesian grid would require $\approx$4$\times$10$^9$ 
cells, orders of magnitude beyond currently available resources.

\begin{table}
\caption{The list of grids}
\begin{tabular}{lcccc} \\
\hline
grid   & inner     & \multicolumn{2}{c}{number of cells} & size of the\\
name   & boundary  & tangential & radial                 & inner cell \\
\hline
Ls     &   100 pc  & 40         & 67                     & 4.0 pc\\
Le     &    20 pc  & 40         & 88                     & 0.8 pc\\
Ms     &   100 pc  & 80         & 132                    & 2.0 pc\\
Me     &    20 pc  & 80         & 174                    & 0.4 pc\\
Hs     &   100 pc  & 160        & 261                    & 1.0 pc\\
\hline
\end{tabular}
\end{table}

We assume that the gas flow is bisymmetric, so we modeled
half of the plane only, and we imposed
periodic boundary conditions in the azimuthal direction. We considered 
both free flow and reflective boundaries in the radial direction, but 
since the circumnuclear gas morphology may depend on the gas entering 
the grid through the inner boundary, such entering has been prohibited,
i.e. on the inner boundary we allow outflow only. The same inner 
boundary condition is used in all models of Piner \etal (1995), and
Englmaier \& Shlosman (2000).

The gas in our models is initially distributed uniformly over the
grid with a surface density 10 \solmtxt pc$^{-2}$, and it orbits at its 
circular velocity in the initially axisymmetric potential. Here we
approximate the gas in a galaxy consisting of a variety of individual 
clouds as an isothermal single-component fluid. If the temperature 
of this fluid is interpreted statistically as a measure of the
cloud velocity dispersion, as suggested by Englmaier \& Gerhard (1997),
the cell size has to be larger
than the average cloud size to make this approach plausible
--- it is at least 0.4 pc at the inner boundary of the grid Me.
We ran models with the sound speed of 5 or 20 km/s, which corresponds
respectively to the velocity dispersion in
the Galactic disk inside the solar radius, and the vertical
cloud velocity dispersion in the Galactic bulge. 
Thus a 2 Gyr run corresponds to 0.31 and 1.25 sound-crossing
times on our grid respectively, and almost 12 orbits of gas at the
corotation radius of the main bar.

\subsection{Potential and forces for a double bar}
We follow gas flows in the potential of Model 2 constructed in Paper I.
This is a model of a dynamically possible doubly barred galaxy. For gas 
flows in a 
single bar, the mass of the secondary bar is redistributed into the 
axisymmetric bulge. The semi-major axis of the main bar is 6 kpc, while 
that of the secondary bar is 1.2 kpc. The bars rotate rigidly and with
uniform speed about the axis perpendicular to the plane of the gas flow.
Pattern speeds of the main and secondary bars are 36 \kms kpc$^{-1}$ 
and 110 \kms kpc$^{-1}$ respectively, which corresponds to rotation
periods 0.171 Gyr and 0.056 Gyr.
The rotation curve of the system with contributions from its components, 
accompanied by a plot of angular frequencies with pattern speeds of the 
bars superimposed is presented in Fig.3 of Paper I.

The potential of a single bar is steady in the frame rotating with it,
and its forces 
on the fluid in each grid cell can be calculated at the beginning of the 
run, and stored for later use. If another, independently rotating
bar is present, there is no reference frame in which the potential 
is steady. Therefore implementation of an additional bar is equivalent 
to implementing a single bar in any uniformly rotating frame. 
At the start of the run, for each grid cell we
calculate forces $F_0$ from the part of the potential which is 
steady in the reference frame rotating with the main bar (\ie from
the main bar, disk and bulge). We also calculate the force $F_S$ from 
the second bar on a grid corotating with this bar, and having the 
same radial zone spacing, but more nodes in the tangential direction
than our computational grid. At any given time, the code finds the 
relative position of the bars,
and interpolates $F_S$ in the tangential direction with the use of 
splines. The gravitational force from the second bar is added 
to $F_0$ to give the total force at any point $(R,\varphi)$. Runs
for a single bar were performed in various reference frames, but runs
for a doubly barred galaxy are all in the frame rotating with the
primary bar. 

In order to avoid spurious shocks and other features that may 
occur when bars are turned on abruptly, we introduce non-axisymmetric 
components of the potential gradually by forming them out of the bulge.
We keep the sum of the masses of bulge and the bars constant within a given 
radius (10 kpc here),
and also hold the central density constant. The disk parameters do not 
change while the bars are being introduced. At the start, the bars are 
absent, and the force is the sum of disk and bulge components. We then 
introduce the primary bar by allowing it to grow linearly in 
mass until it reaches its full strength at 0.1 Gyr (at about half of its
rotation period), and we analyse the switch-on effects by performing
a variation on this run, in which the bar is introduced 5 times more
slowly. In models with a secondary bar, we either introduce the inner
bar simultaneously with the main one, or we wait till 0.5 Gyr
for the gas flow in the main bar to stabilize, and then let the inner 
bar grow linearly for another 0.05 Gyr.
After both bars have been introduced, the total force $F$ is just the 
sum of $F_0$ and $F_S$.

\subsection{Test problems}
The CMHOG code passed various standard tests including shock tubes,
advection, and the strong shock test suite. The radial outflow, 
cylindrical shock tubes and uniform rotation were also tested in
polar coordinates (see Piner \etal 1995). 

We checked how well the code conserves mass and angular momentum, both
with an imposed radial gas flow in an axisymmetric potential, and for the 
flows in our bar models. For the purely radial gas motion, with no rotation, 
mass is conserved very well. For example, even on lowest resolution grid Ls, 
in the test case of supersonic radial
inflow with a free outflow condition at the inner boundary, the error 
in mass calculated from fluxes is of the order of 10$^{-4}$ of the 
total gas mass on the grid, at the time when about 50\% of gas on the grid 
has passed through the boundary. In the models with a single bar on low 
resolution grid Ls, the mass error is 0.25\% of the total mass at the 
end of the run (2 Gyr). We defined the angular momentum error 
by the discrepancy between the two sides of equation
\begin{equation}
\label{angmom}
\frac{d}{dt} \left( \int_{\rm grid} \bf{r} \times \bf{v} \; dM \right) =
\int_{\rm grid} \bf{r} \times \bf{F} \; dM  \; - \;
\int_{\rm bndry} d \bf{l} \; \bf{f}_L ,
\end{equation}
with $\bf{f}_L$ being the angular momentum flux, $d \bf{l}$ a unit vector 
pointing outward from the grid boundary, $\bf{r} \times \bf{F}$ the 
gravitational torque at position $\bf{r}$, and $\bf{v}$ the gas velocity 
there. This error varies at the end of the test run 
from 0.5 to 0.9\% of the total value, so it is larger than the mass error,
but still small. The mass and angular momentum errors 
for the runs with double bars are similar to the ones quoted above.

We also compared gas flows in a model with a single bar 
calculated in a frame rotating with the bar, and in a stationary 
inertial frame. They show no significant differences, which makes us 
believe that the time-varying potential was implemented properly in 
the code.

\subsection{General remarks about hydrodynamical models built in this work}
About 20 hydrodynamical runs were performed, in potentials
of one or two bars for various grids, gas sound speeds, and initial 
conditions. In this paper, we focus on three setups, with a further one
serving as a comparison model; these are listed in Table 2.
Run S05 was performed in order to check consistency of our results with 
those of Athanassoula (1992) and Piner \etal (1995). The gravitational 
potential in this run differs from Athanassoula's standard model
by the Ferrers index of the bar being $n=2$ here, instead of $n=1$.
The dependence on gas sound speed first reported by Englmaier \& Gerhard 
(1997) was investigated in run S20. The gas flow in a dynamically possible 
doubly barred galaxy was explored on the basis of run D05 with
the low sound speed.

\begin{table}
\caption{The list of physical setups modeled here}
\begin{tabular}{cccl} \\
\hline
setup name & bar            & sound speed    & grids with \\
           &                &   $v_s$        & ending time \\
\hline
S05        & single         &  5 km s$^{-1}$ & Ls,Ms,Hs 2.0 Gyr \\
S20        & single         & 20 km s$^{-1}$ & Me 0.9 Gyr\\
D05        & double         &  5 km s$^{-1}$ & Me 2.0 Gyr\\
N05        & nuclear        &  5 km s$^{-1}$ & Le 2.0 Gyr\\
\hline
\end{tabular}
\end{table}

In all the runs, soon after the bars are fully introduced, gas dynamics
enters a global steady state, \ie it does not show any major structural
changes with time. Features of gas flow in bars presented in the
following sections clearly persist throughout the hydrodynamical runs,
and therefore are part of this global steady state. Nevertheless,
the gas flow is never completely stationary on local scales. 
All our runs show the formation 
of density concentrations, that propagate through the shocks, and create
blobs and smudges. In the density diagrams, they move fast towards 
the shock, get shocked, and fall inwards almost along the shock, 
thickening the density enhancement associated with the shock.
The smudge then splits from the shock, forms an arch overshooting 
the nuclear ring downstream from the shock, and enters the opposite shock.
In run S05 this gas density enhancement propagates through the grid with an 
average period of 88 Myr, about half of the bar rotation period. There is 
no significant change of its amplitude throughout the run, and many
features on the snapshot density diagrams are transient.
We performed a number of variations to run S05
on the low resolution grid Ls rotating at various rates, including
one in an inertial frame, so that the position of the bar on the grid
is no longer fixed. The appearance and periodicity of these features
remained unaffected, so we believe that they are not purely numerical
effects. We also repeated run S05 on the Ms grid with the bar being
introduced more slowly, over 0.5 Gyr or almost 3 rotation periods.
This had no significant effect on the overall gas flow morphology,
but it did weaken the non-stationary propagating features. 

Much more nearly stationary than the density diagrams are the
plots where we display the square of velocity divergence in the gas 
({\sl div}$^2{\bf v}$) for areas with {\sl div} {\bf v} $<0$. Negative 
divergence, or convergent flow indicates shocks. Smudges and blobs do not 
appear in the velocity divergence diagrams, showing that these features
do not disturb the velocity field greatly. 

\begin{figure*}
\resizebox{17.5cm}{!}
{\includegraphics{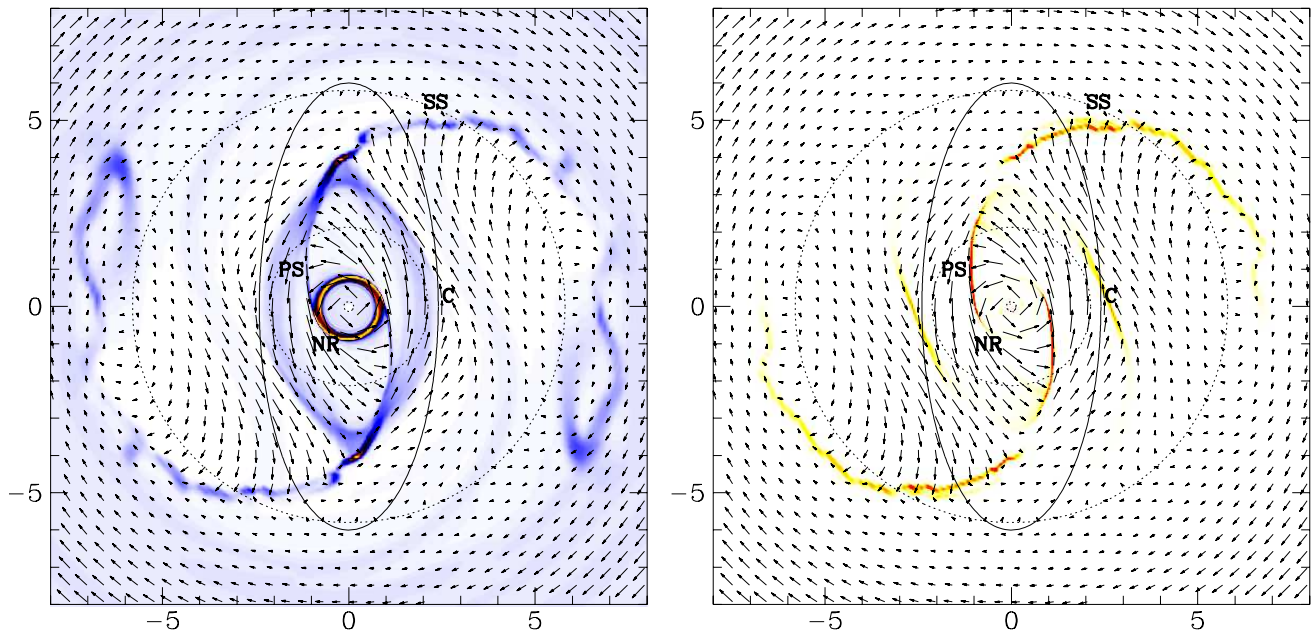}}
\vspace{-15.5cm}
\caption{Density ({\it left}), and ${\sl div}^2\bf v$ ({\it right}) 
snapshots for
gas flow in a single bar (Model S05) on grid Hs made at time 1 Gyr. The bar, 
outlined by the solid line, is vertical. Dotted curves mark resonances in the
azimuthally averaged potential. They are (innermost out): the inner ILR at
0.13 kpc, the outer ILR at 2.14 kpc, and the corotation at 7.78 kpc. Arrows 
mark velocity in the frame rotating with the bar, with length proportional
to the speed. Bare arrowheads mark velocity direction only (in the case of 
small velocities near corotation). Positions of the principal shock 
({\bf PS}), nuclear ring ({\bf NR}), spiral shock ({\bf SS}), and the 
convergence region ({\bf C}) are marked. In the density map, colours follow 
increasing
densities from white through blue, black and red to yellow. The density 
saturates at 200 \solmtxt pc$^{-2}$, and the {\bf PS} is saturated in the 
divergence plot. Units on axes are in kpc.}
\label{s5fig}
\end{figure*}

\section{Fundamental features of flows in a single bar}
General features of gas flow in a single, fast bar with two ILRs can be 
seen in Fig. \ref{s5fig}, which is a snapshot of a run S05 after the flow 
has entered a steady state.
Together with the density distribution, we displayed there the square 
of velocity divergence in the gas ({\sl div}$^2{\bf v}$) for areas where 
{\sl div} 
{\bf v} $<0$, which indicates shocks. The main features of gas flow comply 
well with the basic paradigm described in the Introduction. Two principal 
shocks (marked {\bf PS} in Fig.1) on the leading edge of the bar are almost 
straight, off-centered from the nucleus, and inclined to the bar's major 
axis. The largest values of {\sl div}$^2{\bf v}$ are detected in 
these principal shocks, and the velocity field displayed by arrows shows
that the gas abruptly changes velocity when crossing the shock. The shocked 
gas rapidly falls towards the galaxy center: at the radii where shocks 
are present, gas inflow to the center dominates. In Figure 2, we plot
the gas density, and both components of velocity, for each zone at the radius
1.8 kpc at the same time as in Figure 1. The amplitude of the radial
velocity is large: it varies between $-$140 and $+$140 km/s, more than
half of the circular speed at that radius, and similar to
earlier models. Nevertheless, the density peaks where the radial velocity 
is negative, and on average the inflow dominates.

The inflowing gas settles on near-circular orbits in the nuclear 
ring ({\bf NR}), which in Fig.\ref{s5fig} appears as a highly over-dense 
region at radius about 0.9 kpc,
with no evidence of strong shocks. Under a closer inspection, 
the nuclear ring has the appearance of a tightly wrapped spiral, as
first noticed by Piner \etal (1995). In run S05 the ring extends between 
0.8 and 1.3 kpc, and its position is independent of the resolution.
Contrary to Piner's finding, the ring is not located at the 
maximum of the $\Omega - \kappa/2$ curve, which for Model 2 is at 0.55 kpc. 
The gas motion is mostly circular in the nuclear ring;
there is no inflow beyond it, and gas accumulates there.
Large densities and low shear in the nuclear ring make ideal conditions 
for star formation. If the gas density exceeds the critical value, 
self-gravity should become important, and star formation should occur.

No shocks cross the nuclear ring. There is only a very weak, tightly 
wound sound wave propagating inside the ring. Passing through the
low-density gas there, it compresses it by a factor of 2, but
does not appear to increase the inflow to the center:
for the entire 2 Gyr of the run, the rate of inflow to the center is
below $1 \times 10^{-4}$\solmtxt yr$^{-1}$.
Velocities inside the ring also remain mostly circular: close to 
the inner grid boundary at 0.1 kpc, the radial velocity is uniformly 
zero with largest deviations of 0.3 km/s, while the circular velocity 
is 80--100 km/s. In the next sections, we explore how to increase the inflow 
that is suspended by the nuclear ring.

In addition to the basic features above, Fig.\ref{s5fig} indicates
other generic characteristics of gas flow in a single bar. The principal 
shock is interrupted at about the position of the 4/1 resonance at radius
between 3 and 4 kpc, where no large negative {\sl div}~{\bf v} has been 
detected. At the same position, which is almost on the bar's major axis, the
density plot shows high gas concentration, and the velocity vectors
indicate no strong shear. Thus this is another region that may support
star formation, as seen in NGC 1530 (Downes \etal 1996, Regan \etal 1996)
and in M 94 (Waller \etal 2001). 
Englmaier \& Gerhard (1997) explain the dynamics of this region by
the interaction of gas moving on $x_1$ orbits with that on 4/1 orbits.

Outward of the 4/1 resonance, the principal shock continues as a weaker
spiral shock ({\bf SS}) that reaches the corotation radius, weakens there,
and ends near the Lagrangian points $L_4$ and $L_5$. Gas entering the spiral
shock mainly follows trajectories whose shapes indicate that they may
be related to the banana orbits around these Lagrangian points (compare 
flow lines in \eg Fig.3 and 13 of Athanassoula 1992). As first noted by 
Piner \etal (1995), here the lower-density pre-shock gas is
deeper in the gravitational well than the dense gas past the shock, 
therefore Rayleigh-Taylor instabilities can develop and destroy the 
continuity of the shock.

Gas on banana orbits runs against the shocked gas coming out from the 
principal shock, which creates convergence regions ({\bf C}). This 
convergence funnels gas towards the 4/1 resonance at the end of the principal
shock, which amplifies star formation there.

The existence of the principal and the spiral shocks, as well as of the
convergence regions, is generic for various bar potentials, and 
independent of the sound speed in the ambient gas.

\begin{figure}
\resizebox{8.5cm}{!}
{\includegraphics{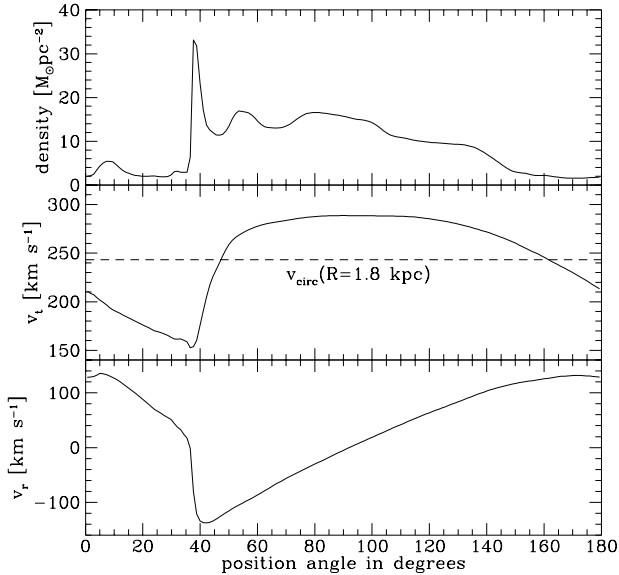}}
\caption{Gas density (upper panel) and both components of velocity (two lower
panels) in gas flow in run S05 at the same time as in Figure 1 for each 
zone at the radius 1.8 kpc. The zones cover 180 degrees starting from the
direction towards the bottom of the figure and going counterclockwise.}
\label{6.5}
\end{figure}

\section{Gas flow inside the central kpc}
The exceptional resolution of a polar grid near its centre prompted us to 
investigate gas flows in the central kiloparsec of a barred galaxy. This
resolution cannot be achieved on a fixed Cartesian grid. Also
models using the SPH method experience a 
rapid drop of the particles onto the center, so any structure detected 
in the inner kiloparsec (Englmaier \& Gerhard 1997, Patsis \& 
Athanassoula 2000) cannot be followed for a long time. The CMHOG code
allowed Piner et al (1995) to analyze and follow the nuclear ring
in exceptional detail. Here, we use this code to investigate
flows in the galaxy center that differ from Piner's model in
properties of the gas and the underlying potential.

\subsection{Dependence on gas sound speed}
Englmaier \& Gerhard (1997) have reported that when the gas in 
barred galaxies is modeled as an isothermal fluid, the 
flow morphology depends on the sound speed. They noticed that 
as the sound speed increases, the off-axis principal shocks
seen in run S05 turn into shocks located almost on the bar's major
axis, with no gas on the $x_2$ orbits. We performed a simulation in the 
same gravitational potential and initial conditions as in 
model S05, but with the gas sound speed increased
to 20 \kms. A snapshot of gas flow in the run S20, which except for the higher
sound speed does not differ from run S05, is 
presented in Fig.3. The principal shock persists, but moves closer to
the bar's major axis, in agreement with models by Englmaier \& 
Gerhard (1997). The outer spiral shock, and the convergent flow
in the trailing side of the bar, retain their shape. 

\begin{figure}
\resizebox{16cm}{!}
{\includegraphics{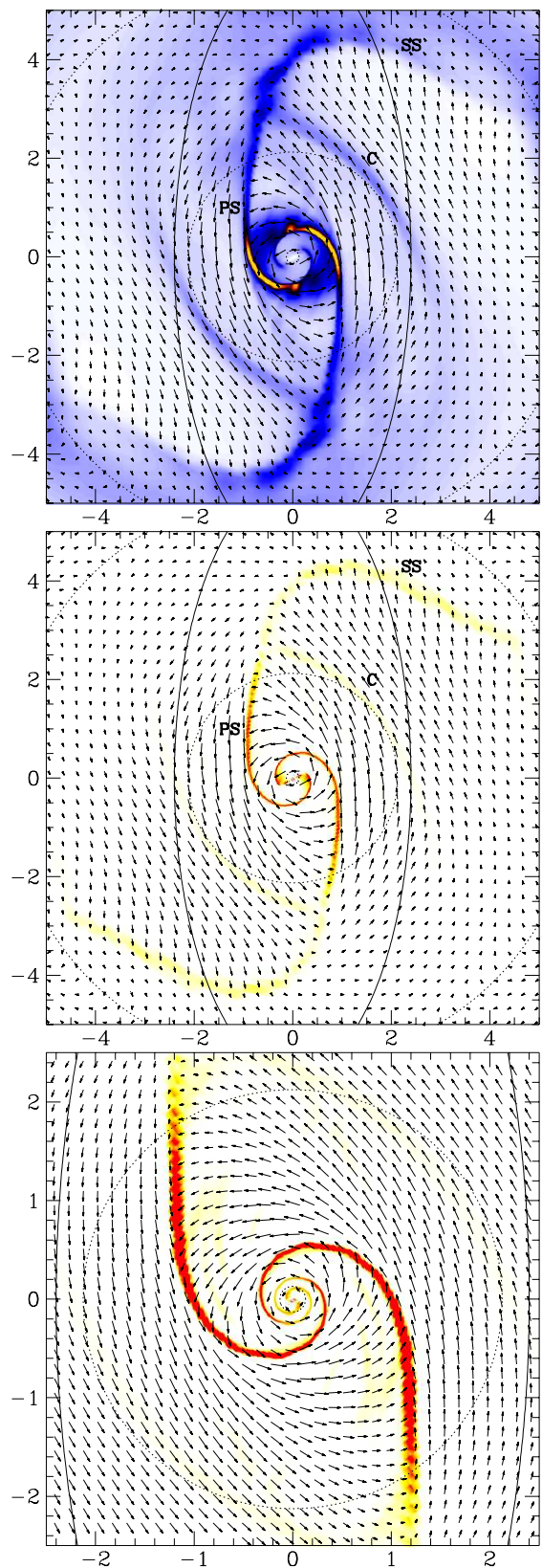}}
\vspace{-1cm}
\caption{Gas flow in the inner regions of model S20 of a single bar on grid 
Me with density (the top panel), and ${\sl div}^2{\bf v}$ (two lower panels), 
presented like in Fig.1. The two upper snapshots are taken at 0.195 Gyr, and 
the lower one at 0.135 Gyr, just before the spiraling shock 
crosses the inner grid boundary. Density saturates at 500 \solmtxt pc$^{-2}$.}
\label{fig4}
\end{figure}

The change of sound speed has the most dramatic effect on the gas 
flow in the region of the nuclear ring. In the low-sound-speed
run S05, the inflowing gas accumulates on the circular nuclear
ring (Fig.1), and the principal shocks do not penetrate into the 
ring. Gas orbits in and inwards the ring are almost circular.
At the high speed of sound (run S20), the nuclear ring turns into 
a nuclear spiral, with pitch angle higher than those of the nuclear ring 
and of the spiral inside it in run S05. The principal shock continues without 
substantial weakening along the nuclear spiral, which allows gas 
inflow all the way to the galaxy center. The inflow through the inner
boundary rapidly increases from $1\times 10^{-4}$\solmtxt yr$^{-1}$ 
at times before the spiral shock reaches the boundary, to $5\times 
10^{-3}$\solmtxt yr$^{-1}$ at time 0.14 Gyr, when the boundary is crossed.
Results after the time when the shock crosses the inner boundary
have to be interpreted with caution. Nevertheless, the inflow
rate keeps increasing, reaching 0.15 \solmtxt yr$^{-1}$ at
the end of the run S20 at 0.9 Gyr. We are currently 
working on following this mode of gas flow properly for longer times, but
even this model, spuriously affected by inner boundary conditions,
shows that the spiral preserves its shape throughout the run. It slightly
tightens and becomes considerably denser: at 0.5 Gyr, the radial density
distribution peaks at 330 \solmtxt/pc$^2$, but by the end of the run,
at 2 Gyr, it is 5 times larger.

At the end of the run, the densest parts of the spiral reside at radii around
0.3 kpc, i.e. well inside the peak of the $\Omega-\kappa/2$ curve,
contrary to the findings of Piner \etal (1995) that the nuclear
ring occurs roughly at the maximum of the $\Omega-\kappa/2$ curve.
The arguments based on the orbital structure, outlined in the Introduction,
indicate that the nuclear ring or spiral needs an ILR in the bar to form. 
Here we find that the ring forms between the inner and the outer ILR, but 
its position depends on gas sound speed, and differs from the maximum of 
the $\Omega-\kappa/2$ curve. Also, there are no grounds for the condition 
postulated by Piner \etal under which the nuclear ring can form. The absence 
of the nuclear ring in their model 8, on which this condition is based, 
results from the potential that does not allow $x_2$ orbits (model 86 in
Athanassoula 1992), and not from the shift in the $\Omega-\kappa/2$ curve.

In the picture presented by Englmaier \& Gerhard (1997), the transition from 
an off-axis principal shock with a nuclear ring to an on-axis one with a 
straight inflow to the center is more or less continuous. By contrast,
in our runs we observe a nuclear spiral at high sound speeds. The study
of on-axis shocks is particularly difficult on polar grids, where special
inner boundary conditions have to be adopted. In general,
we confirm the findings of Englmaier \& Gerhard (1997) that
not only the underlying potential, but also the gas characteristics
influence the morphology of the flow. We extend their conclusion to
the inner kiloparsec of galaxies, where the effect of the gas sound speed
seems to be most dramatic. Inflow to the galaxy center, believed
to be prohibited because of the orbital structure (gas accumulation
on the nuclear ring) becomes possible when velocity dispersion of the
clouds (equivalent to sound speed in the statistical interpretation)
is large enough.
Exploring variation of the gas flow with the sound speed shows that 
many qualitatively different morphologies of the flow can be produced in a 
given external gravitational potential. It is a promising way of 
looking for various modes of gas flow and morphology in central parts 
of galaxies. On the other hand, this shows that gas flows do not 
diagnose the potential uniquely.

\subsection{A dynamically possible doubly-barred model}
Finding particle orbits supporting doubly-barred systems proves 
difficult, since the potential is not steady in any reference frame, and
therefore the Jacobi integral is no longer conserved. This might in
principle mean that orbits are mostly chaotic, but in Paper I we showed
that bars within bars have families of regular particle orbits that
can support them. Further, the orbital structure puts strong constraints on
sizes and rotation rates of the bars. Below we briefly review these 
constraints and their origins in order to show their role in shaping gas flows
in double bars.

In Paper I, we approached the self-consistency issue by searching for 
potentials that generate orbits, with shapes such that the proper distribution
of particles on these orbits could create the density distribution that 
gives rise to the assumed potential. As a first approximation, we
assumed a potential of two rigid bars, rotating at two constant 
incommensurable pattern speeds. This a priori excludes full self-consistency,
because two bars must distort and accelerate as they rotate through 
each other, but proves to be a useful approach to this complicated
dynamical system, which successfully finds orbits supporting double bars.

In a single bar, the family of the $x_1$ orbits, that are aligned with the 
bar, forms the backbone of a steady potential, and particles trapped around
orbits in the  $x_1$ family provide building blocks for the bar. There
is no such simple correspondence in a system consisting of two independently
rotating bars, because the relative period of the bars adds an extra 
frequency to the system, and this frequency is generally incommensurable 
with the orbital periods there. Thus closed periodic orbits are extremely 
rare in double bars, i.e. they form a set of measure zero compared to the case
of a single bar. Nevertheless, it is not the periodic orbits as such, but a 
proper distribution of particles trapped around them, that makes the building 
blocks of the potential. We looked for particles that together remain on a 
closed curve, which returns to its original position, after the two bars have 
come back to the same relative orientation, although the particles 
individually follow various not-closed orbits.
We call such a curve {\it the loop}: stable loops, with 
shapes that follow the two bars in their motion, form the backbone of the 
pulsating potential in a doubly barred system. Particles trapped around such 
loops provide building blocks for a doubly barred galaxy. Fig.4 shows two 
examples of loops in a doubly barred galaxy as the bars rotate through
each other --- the loops change their shape as they follow the bars.
One loop is following the main bar, and the other the 
secondary bar. The position of a randomly chosen particle from the outer
loop is also marked. Note that although loops return to their
original positions once the potential assumes its original shape, the
particle ends up in a different place on the loop than it started from. Its
orbit is not closed. 

\begin{figure}
\resizebox{8.5cm}{!}
{\includegraphics{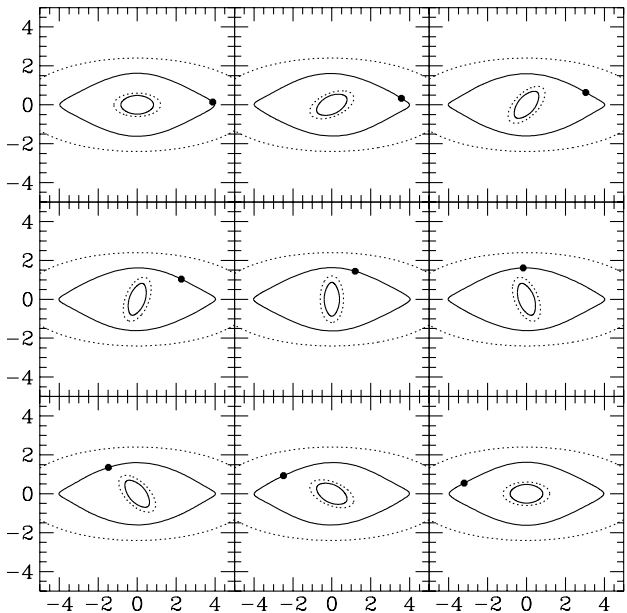}}
\caption{Time evolution of two loops (solid curves) in a doubly barred
potential between two consecutive alignments of bars. Position of one
arbitrary particle on the outer loop is followed and marked. Bars are
outlined by dashed lines. Units on axes are in kpc.}
\label{fig6}
\end{figure}

The concept of a loop proved to be essential to the
realization that doubly barred galaxy potentials allow stable regular 
orbits that can support the shape of both inner and outer bars.
In Paper I, we searched for models in which both bars
have associated families of stable loops, that are elongated with 
the bar, and extend over the bar size. We found that this goal
can be best achieved when the corotation of the inner bar overlaps
with the ILR of the main one (which is the case of resonant coupling
investigated by Tagger \etal 1987). In our nearly self-consistent model 
(Model 2 in Paper I) of a doubly barred galaxy, the most important 
loop families, which serve as the bars' backbones, correspond to the 
$x_1$ and $x_2$ orbits in the single bar. Loops supporting the main 
bar originate from its $x_1$ orbits, and loops supporting the secondary 
bar originate from main bar's $x_2$ orbits. Thus a secondary bar
created in this way can exist only when there is an ILR in the main bar.
Fig.9 of Paper I shows that only inner $x_2$ loops support the secondary 
bar in its motion. Thus the secondary bar ends well inside the ILR of the 
main bar, which also is the corotation of the secondary bar there.
Therefore in the case of such resonant coupling, the secondary bar in Model 2 
of Paper I is slow (it does not extend all the way to its corotation).
This last finding may not be general though, since we assumed simple 
Ferrers' bars in Model 2. Considering bars with more realistic mass 
profiles could weaken the constraints found in Paper I. Nevertheless,
the loop approach allows us to rule out, as being far from self-consistent,
many hypothetical doubly-barred systems. The
requirement of self-consistency puts strict limits on parameters
of acceptable double bars. For this reason it is essential to
examine gas flows in the dynamically possible Model 2.

\subsection{Gas flows in a dynamically-possible double bar}
We were exploring gas flows in a dynamically possible double bar
by performing hydrodynamical runs in the potential of Model 2 
constructed in Paper I. We present the gas flow in only one model of a 
double bar since it is the only near-self-consistent dynamical model 
yet constructed. The complicated dynamics of such systems makes
generating such models a rather time consuming endeavour. Although
we are not certain that our results are generic,
we prefer to avoid investigating gas flows in arbitrary
potentials of double bars, since the results may be misleading, as
we do not know if such double bars can even exist.

Our main run D05 was performed for the gas sound speed 5 \kms. We also 
ran a model N05 in the potential of the secondary bar only, that is 
with the main bar density azimuthally averaged. We first focus on the 
results of the version of run D05, in which the secondary bar was 
introduced 0.5 Gyr after the primary. After that, we comment on the 
effects of simultaneous introduction of the bars.
Run D05 has the same initial and boundary conditions as run S05, and
the general features of the gas flow in this run remain unaffected 
by the presence of the secondary bar: the principal shock, the outer 
spiral shock, and the convergence region retain their shape and position.
Also the appearance of the transient smudges is not affected by the
modified potential, which implies that they are robust features, 
insensitive to model's details.

\begin{figure*}
\resizebox{18cm}{!}
{\includegraphics{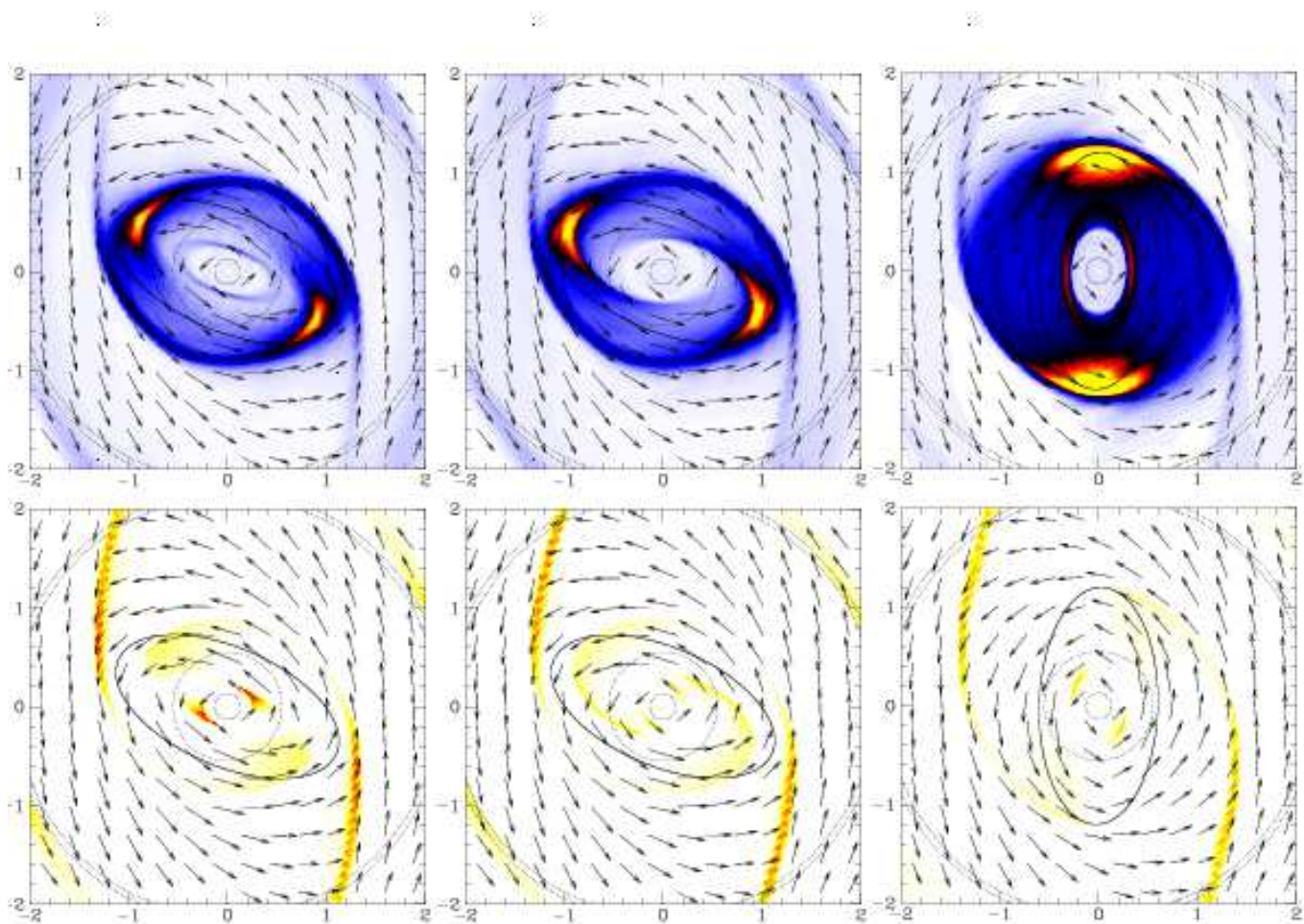}}
\vspace{-13cm}
\caption{Like Fig.1, but for the model D05 of a doubly barred galaxy with 
the secondary bar being introduced at 0.5 Gyr. The three columns are 
close-ups of three different snapshots with density on the top and 
${\sl div}^2\bf v$ on the bottom. The snapshot on the left is taken at 
the time 0.695 Gyr, the one in the middle -- at 0.780 Gyr, and the one
on the right at 2 Gyr. The primary bar is always vertical, and the 
secondary bar is outlined by the solid line. Dotted lines 
indicate (innermost out) positions of the inner ILR of the main bar,  the 
maximum of $\Omega-\kappa/2$ curve at 0.55 kpc, where the secondary bar 
is close to having an ILR, the outer ILR of the main bar, and the 
corotation of the secondary bar at 2.19 kpc. Density saturates at 400 
\solmtxt pc$^{-2}$.}
\label{fig7}
\end{figure*}

Fig.5 shows that, as expected, the secondary bar affects the central 
regions most strongly. The nuclear ring, clear and thin in the singly-barred 
run S05, is widened by the secondary bar (compare Fig.1 to
Fig.5). After a 0.5 Gyr delay, which enables the nuclear ring in the 
primary bar to form, the secondary bar is introduced. Once 
the secondary bar reaches its full strength, at 0.56 Gyr, the inner part 
of the nuclear ring assumes an elliptical shape and begins 
to rotate with the secondary bar. Transitory elliptical rings
(Fig.5 left) and shocks similar to the principal shocks in the 
primary bar (Fig.5 middle) form inside it at various times
without a clear periodicity. Later during the run, the elliptical ring aligned
with the secondary bar is established (Fig.5 right). The semi-minor axis
of this ring is between 0.24 and 0.42 kpc --- it does not extend all the way
down the $x_2$ loop family, which ends at 0.17 kpc (Paper I). 
These late gas morphologies may be
unrealistic, since they correspond to about 30 rotation periods 
of the secondary bar, while in the best N-body models so far the 
secondary bar survives only for several of its own rotation periods. 
On the other hand, our prescription for a nearly self-consistent double
bar presented in Paper I should keep such a system indefinitely long.

If we introduce the secondary bar simultaneously with the primary at the
beginning of the run, the ring associated with the secondary bar forms
very early, already at 0.08 Gyr. It then follows the same evolution as
the central regions in the version with the delayed introduction of the
secondary bar. By the end of the run, the only difference in gas 
distribution is that the ring corotating with the inner bar is about 50\%
denser in the version with the delayed secondary. All the morphology and
kinematics is virtually identical, and the highest-density gas resides in 
`twin peaks' on both ends of the secondary bar throughout both versions
of the run.

Both versions of the run develop an elliptical ring or hole 
inside the secondary bar, which keeps gas away from the galaxy center. 
Once the secondary bar reaches full strength, gas inflow through the inner 
boundary is negligible, about $3\times10^{-5}$ \solmtxt yr$^{-1}$ in run
D05, and gas cannot flow into the center. This contrasts with run S20 in 
a single bar, where some inflow was driven by a spiral 
shock. In the doubly barred potential, gas flow in the flattened nuclear ring 
is smooth, with only weak transient shocks being generated at a special 
relative position of the bars. 

It is easy to understand this modeled flow in terms of the
underlying orbital (loop) structure. In a singly-barred galaxy,
gas follows stable closed periodic orbits, as long as they do not 
intersect or develop cusps. The principal shocks in a single bar
develop because these orbits become cuspy or self-intersecting;
they are offset because of the presence of the anti-aligned $x_2$ 
orbital family. Principal shocks are straight only in a bar that 
extends almost all the way to corotation. Similarly, in the 
potential of double bars, gas stays on or around the stable 
loops. In the dynamically possible Model 2 from Paper I,
loops supporting the secondary bar extend only about half-way 
to its corotation radius, therefore we would not expect straight shocks,
like the principal shocks in the main bar.
Athanassoula (1992) found that even when the ratio of the bar 
size to its Lagrangian radius was lower than 0.71, the principal
dust lanes curl around the bar, and start forming a ring. We do not expect
off-centered shocks within the secondary bar,
because no loops corresponding to the $x_2$ orbits in the secondary
bar were found. An even stronger statement can be made:
in the model we considered, the loops supporting the secondary bar
originate from the $x_2$ orbits in the main bar -- they are 
rather round, with no cusps, so there is no reason for shocks in the gas flow 
to develop. Thus we conclude that the gas flow in some self-consistent doubly 
barred galaxies may lack principal shocks or dust lanes. Nevertheless, this 
conclusion is based on one model only, and a better exploration of 
dynamically possible potentials is needed in order to reach firmer 
conclusions.

We performed an additional run to check to what extent the gas flows 
in the inner part of a doubly barred potential are due to combined 
action of both bars. In particular, we wanted to know, whether there can 
exist principal dust lanes in the small bar when the primary bar is absent. 
Run N05 has the same initial and boundary conditions 
as run D05, but here the primary bar is not included;
we introduce the secondary bar alone. We found the gas flows
in the inner part of the galaxy to be very similar to those in run D05
throughout the evolutionary time. Except for the absence of the nuclear 
ring, and lower peak densities in the gas (both evidently caused by the 
action of the primary bar, which is not present here), the gas flow in the 
secondary bar retains all the features of run D05.

\section{Discussion}

\subsection{General Remarks}
Our high-resolution simulations of gas flow
in barred galaxies allowed us to resolve and to follow features
of the flow with exceptional detail. In Section 2.4 we noticed that
the flow in a single bar never becomes fully stationary, and gas 
condensations propagate with unattenuated amplitude. If the bar is
introduced more slowly, these features are weaker. We may speculate
that irregular gas morphologies are more likely to appear in bars that formed
fast, for instance as a result of interaction of two galaxies passing
close to each other. We may see evidence of this effect in the interaction
between NGC 1410 and NGC 1409, which perturbed the first galaxy, and
probably led to star formation in its centre and at the 4/1 resonance 
along the bar. Blobs and smudges appear to connect the 4/1 resonances 
on the two opposite sides of the bar (Keel 2000).

\subsection{Density-wave theory for non-self-gravitating gas in an imposed 
potential}
Both runs in a single bar presented here display a spiral feature
extending almost to the galaxy centre, but its appearance varies
strongly with the gas sound speed. In the linear approximation, 
this spiral can be interpreted in terms of the gas density 
waves adopted by Englmaier \& Shlosman (2000). In a
standard density-wave theory for gaseous disks, the dispersion 
relation between the radial wavenumber $k$ and the frequency 
$\omega$ is derived from the linearized continuity and Euler's
equations (see \eg Binney \& Tremaine 1987, pp.355-359, hereafter BT).
If these equations describe self-gravitating gaseous disks, then 
potential is coupled to density, and the linearized equations are 
homogeneous, with wave solution governed by the dispersion relation 
given by eq.(6-40) in BT for the limit of a tightly wound spiral.
 
Englmaier \& Shlosman note that, 
unlike the standard density-wave theory, our case is the case 
of non-self-gravitating gas in an imposed potential. In this case, the 
potential term makes the linearized Euler's equations (BT eq.6-26)
inhomogeneous, with non-wave solutions. A homogeneous counterpart of
these equations, with no gravity term $\Phi_a$, can be solved
for  a tightly wound spiral
yielding the same dispersion relation as eq.(6-40) in BT,
but with the gravitational constant $G=0$. Since the wave frequency
of an $m$-fold spiral is related to its pattern speed $\Omega_p$
by $\Omega_p = \omega / m$, the dispersion relation can be rewritten
as
\begin{equation}
(\Omega + \kappa/m - \Omega_p) (\Omega - \kappa/m - \Omega_p) 
= \frac{k^2v_s^2}{m^2},
\end{equation}
where $\Omega$ and $\kappa$ are the angular and epicyclic frequencies
respectively, and $v_s$ is the gas sound speed.
Thus, for a given $m$, $kv_s$ is a function of the imposed potential.
Since the pitch angle $i$ of the spiral at a given radius $R$ is defined by 
$\tan i = |m/kR|$, in the tightly wound limit this angle is proportional
to the sound speed $v_s$. Differentiating the dispersion relation (4)
with respect to $k$ shows that the group velocity of
the wave ($v_g = \partial \omega (k,R) / \partial k$) can be written
as the sound speed times a function of the potential. Several other 
interesting conclusions for non-self-gravitating spirals in realistic 
galactic potentials can be drawn from the dispersion relation (4). 
This relation prohibits the single-arm spirals in the inner part of 
the potential, since they can propagate only for 
$\Omega_p > \Omega + \kappa$ or $\Omega_p < \Omega - \kappa$. The $m=2$ 
spiral is confined to within its own ILR, or outside its own OLR (Outer 
Lindblad Resonance), as originally noticed by Englmaier \& Shlosman 
(2000). Higher-$m$ spirals are confined inside their $\Omega - \kappa/m$
resonance or outside their $\Omega + \kappa/m$ one.

The inhomogeneous term in the Euler's equations comes from the external 
potential of the bar, and in the linearized form can be written as 
\begin{equation}
\Phi_1 = \Phi_a(R) \cos(2\phi-2\Omega_b),
\end{equation}
where $\Omega_b$ is the bar pattern speed. Although this term still bears a 
wave-like form, it no longer describes a tightly wound 
spiral, and one cannot use the approximations that led to the dispersion 
relation above. Nevertheless, solution of the linearized continuity and 
Euler's equations will return the coefficients of wave harmonics as functions 
of the imposed potential $\Phi_a$. Only the $m=2$ and $\omega=2\Omega_b$
modes are driven by this form of the potential. In principle, other modes
can propagate in the gas as well, but they will be damped numerically in
the models, or by viscosity in real galaxies, and only the bar-driven modes
will survive.

Therefore, in non-self-gravitating gas various spiral modes can propagate
inside their own ILRs, but they will fade away unless something drives them.
The only driven modes are those of $m=2$ and $\omega=2\Omega_b$, that is
of a 2-arm spiral whose pattern is locked with the bar. This double-arm
spiral will not extend beyond the ILR of the bar.

These conclusions are supported by our recent simulations of gas flow
in very weak bars (Maciejewski 2001). The spiral pattern in the gas
does not extend beyond the outer ILR, or within the inner ILR of the
potential. Nevertheless, if we add a central black hole that removes 
the inner ILR, the spiral extends all the way to the inner grid boundary.

In the low sound speed run S05,
the spiral is weak and rather tightly wound. It creates at most
a factor of two density enhancement over the interarm region, and 
${\sl div}^2\bf v$ in the spiral is ten times smaller than in the principal 
shock. This spiral can be well understood in terms of the gas density 
waves. The nuclear spiral in this run terminates around the radius of 0.19
kpc, just outside the azimuthally averaged inner ILR of Model 2,
as postulated by the linear theory above.

On the other hand, in the high sound-speed run S20, the 
strong, loosely-wound spiral is a direct continuation of the principal 
shock -- there is no nuclear ring to interrupt it. This generally
agrees with the density wave theory, which postulates a higher pitch
angle for a higher sound speed. Nevertheless, at 0.13 Gyr, just 
before the spiral collapses onto the center, the shock strength at
the radius of 0.08 kpc, as measured by ${\sl div}^2\bf v$, is
only twice smaller than the average in the principal shock. The
arm/interarm density ratio is as high as 10. This nuclear spiral
reaches 0.05 kpc, almost three times closer to the nucleus than
the position of the inner ILR. This nuclear spiral is definitely beyond 
the linear regime explored in the density-wave theory: there, sound waves 
cannot penetrate inward past the inner ILR, while the main 
shock in our model extends very close to the centre, crossing the inner ILR.

\subsection{Relating observed and modeled features}
In-spiraling shocks may in fact operate in real galaxies. Both
Regan \& Mulchaey (1999) and Martini \& Pogge (1999) report
an unexpectedly high frequency of nuclear spiral patterns in their
samples, of 12 and 24 Seyfert 2 galaxies respectively. The patterns were 
detected by comparing visual with infrared HST images in combined
color maps, and they  indicate strong dust extinction in the spiral
arms. Estimates of gas density based on this extinction lead Martini 
\& Pogge to the conclusion that these spirals are caused by shocks in 
nuclear disks. The linear theory of gas density waves assumes low 
arm-interarm density contrast, with extinction smaller than 0.1 mag
(Englmaier \& Shlosman 2000), and no extensive star formation. The observed 
dust features can have $A_V \sim 0.5-1.5$ mag higher than the interarm 
region (Quillen \etal 1999), and the nuclear spirals show star formation
(see \eg Phillips \etal 1996; Laine \etal 1999).

The abovementioned two surveys of Seyfert 2 galaxy centers
aimed to resolve the issue of the AGN feeding mechanism. Regan \& 
Mulchaey (1999) searched for evidence of secondary stellar bars by
looking for dust extinction in a form of straight lanes, analogous to the
principal shocks in large-scale bars, which should be 
clearly seen in the color maps. Half of their sample shows
nuclear spirals, and they report only three galaxies that have morphology 
consistent with gas flows in the bars. Out of those three galaxies, 
dust lanes in one (NGC 5347) 
are just continuation of the principal dust lanes in the main bar, 
another one (NGC 7743) shows a single straight dust lane on one side
and circular dust morphology on the other. The only galaxy in that
sample that is known to be doubly barred (NGC 3081) 
differs from the rest of the sample by virtue of having a central 
ring flattened along the secondary bar, and no straight dust lanes.
This is just what our model D05
for a dynamically possible doubly-barred potential predicts.
The outer parts of the NGC 3081 ring are blue -- our simulations show that 
there is no shocks or strong shear there, so this is a possible
site for star formation.
Similarly, out of four best candidates for nuclear bars in the sample
of Martini \& Pogge (1999), two (NGC 5929 and Mrk 270) have no
large-scale bar, out of which one (Mrk 270) shows straight dust 
lanes in the centre. 
A set of straight dust lanes in Mrk 471 is probably continuation
from the large-scale bar. The case of the fourth candidate (Mrk 573)
is complicated because of interaction between the disk and the 
ionization cone (Quillen \etal 1999).

Regan \& Mulchaey take the lack of straight nuclear dust lanes to be
evidence against the presence of secondary bars, and so they conclude 
that secondary bars are not a viable mechanism for feeding the active 
center of a Seyfert galaxy. Although our models confirm that the 
secondary bars may be inefficient in amplifying gas inflow to the 
galaxy centers, we also find that secondary bars do not have 
to reveal themselves through straight dust lanes --- 
gas flows induced by them can be different from those caused by 
single (main) bars.

Finally, it is worth re-examining how the modeled features of gas flow in 
barred galaxies correspond with the best quality observations.
Two major discrepancies arise here. First, we expect that strong shear
in the principal shocks should prevent star formation there, while
observations of early-type galaxies sometimes show strong H$\alpha$
emission along the dust lanes (NGC 986, NGC 7582; Hameed \& Devereux
1999). Second, observations of radio polarization in barred galaxies
(Beck \etal 1999) show the regular magnetic field well aligned with
the numerically modeled gas streamlines, while the depolarization 
region, expected around the principal shock, is shifted from it
in the upstream direction by up to 0.9 kpc. Together, these two
observations indicate that we are still far from the proper 
description of the postshock region in barred galaxies.
Beck's observations of the magnetic field may also provide us with 
insight to the central galactic kiloparsec, where the field seems not to 
be aligned with the expected gas flow.

\section{Conclusions}
Using a high-accuracy hydrodynamical code, we investigated 
various modes of gas flow in a barred galaxy in an imposed
gravitational potential of a single or double bar.
We find that both circular motion of gas in the nuclear ring, and 
spiral shocks extending almost all the way to the galaxy center, are 
possible in a singly barred galaxy with an Inner Lindblad Resonance. 
At high sound speed, gas inflow to the center can proceed through the 
nuclear in-spiraling shock. At low sound speed the inflow is prevented by the
nuclear ring.

It has been commonly assumed that the secondary bar embedded within a 
large-scale bar will generate gas flows, that are simply a small-scale 
analogue of those in the main bar (Regan \& 
Mulchaey 1999). Models of arbitrarily chosen double bars (Maciejewski
\& Sparke 1997, Athanassoula 2000) are in accordance with this assumption, 
and they even predict periodic feeding of the nucleus by the inflow 
in the secondary bar. On the other hand, the loop approach that we 
developed in Paper I made clear that the potential adopted 
by Maciejewski \& Sparke (1997) is unrealistic (Model 1 in Paper I).

Although we have constructed only one dynamically possible double-bar
model (Model 2 in Paper I), and we cannot claim generality of our results,
we argued that in this class of simple potentials, the orbital structure 
confines the secondary bar well within its corotation. The gas flow in this 
bar follows orbits that do not intersect or develop cusps. We have shown here 
that such a bar can bring gas closer to the galactic center, but it does 
not create stationary shocks in the gas flow similar to the principal
shocks in the fast-rotating main bar, neither does it enhance the 
gas inflow to the center. This conclusion, supported
by observations of Seyfert galaxy centers, goes against the
common expectation that secondary bars should activate a stronger
inflow in barred galaxies. We find that the secondary 
bar can prevent, rather than enhance, the inflow.

\section*{Acknowledgments}

We would like to thank the referee, Jerry Sellwood, for useful remarks that
lead us to clarify and improve this paper.
WM would like to thank Mordecai-Mark Mac Low for the idea of examining
shocks in the {\sl div}$^2{\bf v}$ diagrams, Panos Patsis and Peter Erwin for 
useful discussions, and James Binney for comments on this manuscript.
This research was partially supported by NSF Extragalactic 
Program grant AST93-20403. WM acknowledges a postdoctoral fellowship from
the Max-Planck-Institut f\"ur Astronomie, Heidelberg, where part of the
numerical simulations were performed.

\end{document}